\documentclass{aastex}
\usepackage{spr-astr-addons}
\usepackage{url}

\DeclareGraphicsExtensions{.eps,.bmp,.wmf,.jpg,.pdf}

\RequirePackage{color}

\begin{document}

\title{Slow-roll inflation from massive vector fields non-minimally coupled to gravity}
\shorttitle{Slow-roll inflation from massive vector fields}
\shortauthors{A.~Oliveros}

\author{A. Oliveros}
\affil{Programa de F\'isica, Universidad del Atl\'antico, Km 7 antigua v\'ia Puerto Colombia, Barranquilla, Colombia}

\email{emaila}

\begin{abstract}
In this work we study  slow-roll inflation for a  vector-tensor model with  massive vector fields non-minimally coupled to gravity. The model under consideration has  arbitrary parameters for each geometrical coupling. Taking into account a spatially flat FRW type universe and a general vector fields (with temporal and spatial components), we get the general expressions for equation of motion and the  total energy momentum tensor.  In this scenario, the isotropy of expansion  is guaranteed considering a triplet of orthogonal vector fields, but the effective mass of the vector field is of the order of the Hubble scale  and  the inflationary regime is difficult to realize with this model. However, for suitable values (or constraints) of model parameters, it is possible to overcome this issue. In this sense, two cases were analyzed. In the first case, a regime with slow-roll inflation was obtained, and  for the second case the vector field  behaves as a constant, and it drives a quasi de Sitter expansion, hence that slow-roll takes place and inflation occurs.
\end{abstract}



\section{Introduction}

The Hot Big Bang model of the universe has many long-standing problems, for example: the horizon, the flatness, and the monopole problems, among
others. The inflationary paradigm (Inflation) was introduced to solve these problems [\cite{guth, albrecht, linde}]. Inflation is a period of accelerated expansion of the universe in early times and in this scenario, the space grows exponentially (or quasi-exponentially) fast for a fraction of a second after the Big Bang. Furthermore, Inflation  can explain the current temperature
fluctuations observed  in CMB spectrum and the formation of the large scale structures  in the universe [\cite{chibisov, hawking, guth2}], and also predicts a nearly scale invariant primordial power spectrum [\cite{lyth, lidsey}].\\

\noindent Many different models of inflation have been proposed and studied in the literature. In
these models inflation is generically driven by the coupling of one or more scalar fields to
gravity (the so-called inflaton field), and the dynamics during inflation is such that generically the potential energy of the
fields dominate over their kinetic term and the potential is flat enough to ensure the so-called
slow-roll inflation (for more details about this topic see  [\cite{linde2}] and   [\cite{liddle}]). Moreover, the models of inflation with scalar fields are successful  because others models with
higher spin fields generically induce a spatial anisotropy and the effective masses of such fields are usually of the order of the Hubble scale and the slow-roll inflation does not occur [\cite{ford, burd, chiba}]. Besides of scalar fields models of inflation, in recent years some authors have considered another alternative, suggesting the possibility that the inflation is driven  by  massive vector fields
[\cite{golovnev, setare, darabi, soda, koh}]. The vector fields,  also  have been considered as another alternative to solve the dark energy problem [\cite{kiselev, wei, mota, maroto1, maroto2, harko, oliveros}]. In  the reference [\cite{golovnev}] the authors propose a scenario where inflation is driven by non-minimally coupled massive vector fields and it is shown that in an FRW type universe these
fields behave in precisely the same way as a massive minimally coupled scalar field. In addition, the problem with the spatial anisotropy  is solved considering a triplet of orthogonal vector fields or for the expense of N randomly oriented vector fields. In this sense, we propose
a general vector-tensor model of inflation which has  extra  terms coupled to vector fields (the Ricci tensor, Ricci scalar and Gauss Bonnet invariant). This proposal is more general than the studied  by [\cite{golovnev}].\\

\noindent  This paper it is organized as follows: in section \ref{sec_model} we introduce a vector-tensor model of inflation with non-minimally  coupled terms to  massive vector fields and the  corresponding field equations are obtained. In section \ref{sec_cosmoEvol}, we consider a flat FRW type universe and a vector field with temporal and spatial components, and from these considerations  general expressions for equation of motion and the  total energy momentum tensor are obtained. Also, in this section we consider suitable values (or constraints) of model parameters, and the respective analysis is performed.  Finally, some conclusions are exposed in section \ref{sec_concs}.

\section{The model}
\label{sec_model} The  action for the vector-tensor model of inflation considered in this work has the following form
\begin{equation}\label{eq1}
\begin{aligned}
S = &\int d^4x \sqrt{-g}\Big[-\frac{R}{2} 
    -\frac{1}{4}F_{\mu \nu} F^{\mu \nu}  
    -\frac{\eta}{2} R_{\mu \nu}A^{\mu}A^{\nu}\\ 
    &+\frac{1}{2}\left(m^2+\omega R  
    +\xi \mathcal{G}\right) A_{\mu}A^{\mu}\Big],
\end{aligned}
\end{equation}
where $F_{\mu \nu}=\partial_{\mu}A_{\nu}-\partial_{\nu}A_{\mu}$, $R_{\mu\nu}$ and $R$ are the Ricci tensor and the Ricci scalar, respectively, $\mathcal{G}=R^2-4R_{\mu\nu}R^{\mu\nu}+R_{\mu\nu\alpha\beta}R^{\mu\nu\alpha\beta}$ is the topological Gauss-Bonnet invariant (GB), and free coupling constants ($\eta$, $\omega$, $\xi$) for each non-minimal coupling term are included. $\omega$ and $\eta$ are dimensionless, but the dimensions of $\xi$ are $M^{-2}$. The action Eq. (\ref{eq1}) satisfies three very important conditions: (1) the Lagrangian density is a four-scalar; (2) the resulting theory is metric and (3) there are no higher than second derivatives in the resulting field equations [\cite{mota}].
In addition, to avoid a possible ghost instability of the longitudinal component of the massive $U(1)$ vector field in the slow-roll regime,the following condition needs to be satisfied [\cite{peloso, germani}]
\begin{equation}\label{con1}
\frac{1}{2}\left(m^2+\omega R +\xi \mathcal{G}-\eta R_{\mu\nu}g^{\mu\nu}\right)>0.
\end{equation}
Nevertheless, in order to  decide whether or not this vector model of inflation is perturbatively unstable, the full gravitational and field theory perturbation analysis must be performed. The ghost instability problem will not be considered in this work.\\

\noindent The Eq. (\ref{eq1}) represents a generalization of the model considered in the reference [\cite{golovnev}], where inflation is driven by non-minimally coupled massive vector fields, but the couplings with GB  invariant and the Ricci tensor are not considered ($\xi = 0$ and $\eta=0$). A similar model was studied by [\cite{oliveros}] in the dark energy context, but without explicit potential terms. A model with vector field coupled to GB invariant and  the Ricci scalar in a Bianchi type-I universe was proposed by [\cite{mota}].\\

\noindent The variation of the action (\ref{eq1}) with respect to the metric tensor $g_{\mu\nu}$ gives the field equations
\begin{equation}\label{eq2}
R_{\mu \nu} - \frac{1}{2}g_{\mu \nu}R = \kappa^2 T_{\mu \nu},
\end{equation}
where $\kappa^2=8\pi G=M_p^{-2}$ and $T_{\mu \nu}$ is the energy momentum tensor for the vector field, and which the following form:
\begin{equation}\label{eq3}
T_{\mu \nu} = T^{(1)}_{\mu \nu}+T^{(2)}_{\mu \nu}+T^{(3)}_{\mu \nu}+T^{(4)}_{\mu \nu}+T^{(5)}_{\mu\nu}.
\end{equation}
Each term is given by the following expressions\linebreak (Eqs.~(\ref{eq4})-(\ref{eq8})):
\begin{equation}\label{eq4}
T^{(1)}_{\mu \nu} = F_{\mu \beta} F^{\,\,\,\, \beta}_{\nu} -\frac{1}{4} g_{\mu \nu} F_{\alpha \beta} F^{\alpha \beta},
\end{equation}

\begin{equation}\label{eq5}
\begin{aligned}
T^{(2)}_{\mu \nu} =&\frac{\eta}{2} \Bigl( g_{\mu \nu} \Bigl[R_{\alpha \beta} A^{\alpha} A^{\beta} - \nabla_{\alpha} \nabla_{\beta} (A^{\alpha} A^{\beta}) \Bigr]\\
 &- \square(A_{\mu}A_{\nu}) \,+ 2\nabla_{\beta} \nabla_{(\mu} (A_{\nu)} A^{\beta})\\
&- 4 R_{\beta (\mu } A_{\nu)} A^{\beta} \Bigr),
\end{aligned}
\end{equation}

\begin{equation}\label{eq6}
\begin{aligned}
T^{(3)}_{\mu \nu}& = -\frac{1}{2}\xi \Bigl( 8\Bigl[ R^{\,\,\, \alpha \beta}_{\mu \,\,\,\,\, \nu} \nabla_{\alpha} \nabla_{\beta} (\phi)+ R_{\mu \nu} \square \phi\\
&- 2\nabla_{\beta} \nabla_{(\mu} (\phi)R_{\nu)}^{\,\,\,\beta} + \frac{1}{2}R \nabla_{\mu} \nabla_{\nu} \phi \Bigr]\\
&\,+  4\Bigl[ 2 R^{\alpha \beta} \nabla_{\alpha} \nabla_{\beta} (\phi) -R \square(\phi) \Bigr] g_{\mu \nu} -2\mathcal{G} A_{\mu} A_{\nu} \Bigl),
\end{aligned}
\end{equation}

\begin{equation}\label{eq7}
\begin{aligned}
T^{(4)}_{\mu \nu} = & \omega\Bigl[ R A_{\mu}A_{\nu} + \Bigl(R_{\mu \nu} - \frac{1}{2} g_{\mu \nu} R \Bigr) \phi + g_{\mu \nu} \square (\phi)\\
&- \nabla_{(\mu} \nabla_{\nu)} (\phi)  \Bigl],
\end{aligned}
\end{equation}

\noindent and 
\begin{equation}\label{eq8}
T^{(5)}_{\mu \nu} = m^2 A_{\mu} A_{\nu} - \frac{1}{2}m^2 A_{\lambda} A^{\lambda}g_{\mu \nu} ,
\end{equation}
where $\phi=A_{\alpha}A^{\alpha}$, is an invariant scalar. For more details about the variation of the Gauss-Bonnet term, see
the reference [\cite{odintsov}].\\

\noindent On the other hand, the variation of the action with respect to $A_{\mu}$, gives the equation of  motion 
\begin{equation}\label{eq9}
-\nabla_{\mu} F^{\mu \nu}+ \eta R^{\nu}_{\,\,\mu} A^{\mu} - (m^2+\omega R+\xi \mathcal{G})A^{\nu}=  0 .
\end{equation}

\section{Cosmological analysis}
\label{sec_cosmoEvol}
\noindent In this section we carry out the analysis related to the early cosmological evolution of the universe generated by the model described in the previous section,  taking into account a possible slow-roll inflation regime. First, we consider the flat Friedmann-Robertson-Walker (FRW) metric for a homogeneous and isotropic universe, which is given by
\begin{equation}\label{eq10}
ds^2=dt^2-a(t)^2\sum_{i=1}^{3}{(dx_i)^2},
\end{equation}
where $a(t)$ is the scale factor. Also, for the next calculations, we will regard that the vector field has temporal and spatial components, i.~e.~$A^{\mu}=(A^0(t,\vec{r}),\vec{A}(t,\vec{r}))$.\\

\noindent Using the FRW metric in Eq.~(\ref{eq9})  we obtain the following equations:
\begin{equation}\label{eq11}
\begin{aligned}
&-\frac{1}{a^2}\Delta A_0+\frac{1}{a^2}\partial_i \dot{A}_i-3\eta (\dot{H}+H^2)A_0\\
&+[m^2-6\omega(\dot{H}+2H^2)+24\xi H^2(\dot{H}+H^2)]A_0=0,
\end{aligned}
\end{equation}
\begin{equation}\label{eq12}
\begin{aligned}
&\ddot{A}_i-\partial_i \dot{A}_0+H\dot{A}_i-H\partial_i A_0-\frac{1}{a^2}\Delta A_i\\
&+\frac{1}{a^2}\partial_i(\partial_k A_k)+\eta(\dot{H}+3H^2)A_i\\
&+[m^2-6\omega(\dot{H}+2H^2)+24\xi H^2(\dot{H}+H^2)]A_i=0,
\end{aligned}
\end{equation}
where $\partial_i\equiv \partial/\partial x^i$, a dot denotes a derivative with respect to the physical time $t$
and we assume the summation over repeated spatial indices. One of the quantities
which characterizes the strength of the vector field in coordinate independent way is the scalar [\cite{golovnev}] 
\begin{equation}\label{eq13}
\phi=A_{\alpha}A^{\alpha}=A_0^2-\frac{1}{a^2}A_iA_i,
\end{equation}
from which it is possible to introduce a new variable $B_i\equiv A_i/a^2$ instead of $A_i$. Considering that the vector field is quasi-homogeneous
($\partial_i A_{\alpha}=0$),  it is evident from the Eq. (\ref{eq11}) that
\begin{equation}\label{eq14}
A_0=0,
\end{equation}
and  Eq.  (\ref{eq12}) it can be rewritten in terms of the field strength $B_i$, then
\begin{equation}\label{eq15}
\begin{aligned}
&\ddot{B}_i+3H\dot{B}_i+[m^2+(1+\eta-6\omega)\dot{H}\\
&+(2+3\eta-12\omega)H^2+24\xi H^2(\dot{H}+H^2)]B_i=0.
\end{aligned}
\end{equation}
We can see that, if $\eta=0$, $\omega = 1/6$ and $\xi=0$, then the equation of motion for $B_i$  is reduced to that of
minimally coupled massive scalar fields [\cite{golovnev}]. Now, it is necessary to check if in this scenario the vector field could generate an 
early accelerating phase (inflation). From Eq. (\ref{eq15}), we can define the effective mass of the vector field as
\begin{equation}\label{eq16}
\begin{aligned}
 m^2_{\rm{eff}}\equiv & m^2+(1+\eta-6\omega)\dot{H}+(2+3\eta-12\omega)H^2\\
&+24\xi H^2(\dot{H}+H^2),
\end{aligned}
\end{equation}
In the reference [\cite{koh2}] it is shown that using a timelike vector field, the inflationary regime is difficult to realize, since the effective mass of the vector field is order of the Hubble scale (it is required that $m^2_{\rm{eff}}\ll H^2$) and the slow-roll conditions could not be fulfilled. We can see that the geometrical couplings $R_{\mu \nu}A^{\mu}A^{\nu}$,  $\omega R A_{\mu}A^{\mu}$
and $\xi \mathcal{G} A_{\mu}A^{\mu}$ present in our model produce  a large effective mass (see Eq. (\ref{eq16})), and therefore the inflationary regime is difficult to realize with this model. But, for suitable values (or constraints) of model parameters it is possible to overcome this issue. These restrictions will be taken under consideration in the next section.\\

\noindent From Eqs. (\ref{eq3})-(\ref{eq8}) and using the FRW metric given by Eq. (\ref{eq10}), the components of the energy momentum tensor (Eq. (\ref{eq3})) can be written as
\begin{equation}\label{eq17}
\begin{aligned}
T_0^0=&\frac{1}{2}(\dot{B}_k^2+m^2B_k^2)+(1+\eta-6\omega)H\dot{B}_kB_k\\
&+\Big(\frac{1}{2}-3\omega\Big)H^2B_k^2+24\xi H^3\dot{B}_kB_k,
\end{aligned}
\end{equation}
and
\begin{equation}\label{eq18}
\begin{aligned}
T_j^i&=\Big[\frac{1}{2}\big[(-1-4\omega+16\xi H^2)\dot{B}_k^2+(m^2-(1+6\omega)H^2\\
&-4\omega\dot{H})B_k^2\big]-(1+\eta+\omega-16\xi (\dot{H}+H^2))H\dot{B}_kB_k\\
&-2(\omega-4\xi H^2)\ddot{B}_kB_k\Big]\delta_j^i+(1+\eta)\dot{B}_i\dot{B}_j\\
&+\Big(1+\frac{3}{2}\eta\Big)H(\dot{B}_iB_j+\dot{B}_jB_i)\\
&+[-m^2+(1-3\eta+12\omega-24\xi H^2)H^2-(\eta-6\omega\\
&+24\xi H^2)\dot{H}]B_iB_j+\frac{\eta}{2}(\ddot{B}_iB_j+\ddot{B}_jB_i),
\end{aligned}
\end{equation}
where we have considered that the vector field is homogeneous and the summation over index $k$ is assumed.
For $\eta=0$, $\omega = 1/6$ and $\xi=0$ the Eqs. (\ref{eq17})
and (\ref{eq18}) are reduced to those reported by [\cite{golovnev}] (note that to eliminate the second order
terms $\ddot{B}_kB_k$, $\ddot{B}_iB_j$ and $\ddot{B}_jB_i$ it is necessary to use the equation of motion Eq. (\ref{eq15})).
The Eq. (\ref{eq17}) is
similar to that reported in the literature for scalar-tensor models of dark energy with Gauss-Bonnet and non-minimal couplings [\cite{odintsov, granda}]. It is clear that the spatial part of the energy momentum tensor given by Eq. (\ref{eq18}) contains non-diagonal components, 
therefore, the spatial isotropy is broken. To solve this problem, we will use the formalism  presented in the reference [\cite{golovnev}], namely,   several fields will be considered simultaneously.\\

\noindent We consider a triplet of mutually orthogonal vector fields $B_i^{(a)}$, with the same magnitude $|B|$ each.
The orthogonality of the vector fields is given by
\begin{equation}\label{eq19}
\sum_iB_i^{(a)}B_i^{(b)}=|B|^2\delta_b^a,
\end{equation}
and from this result it is clear that 
\begin{equation}\label{eq20}
\sum_iB_i^{(a)}B_j^{(a)}=|B|^2\delta_j^i.
\end{equation}
From these relations, the Eqs. (\ref{eq17}) and (\ref{eq18}) take the following form
\begin{equation}\label{eq21}
\begin{aligned}
T_0^0=&\rho=\frac{3}{2}(\dot{B}_k^2+m^2B_k^2)+3(1+\eta-6\omega)H\dot{B}_kB_k\\
&+3\Big(\frac{1}{2}-3\omega\Big)H^2B_k^2+72\xi H^3\dot{B}_kB_k,
\end{aligned}
\end{equation}
and
\begin{equation}\label{eq22}
\begin{aligned}
T_j^i=&-p\delta_j^i=\Big\{\frac{1}{2}\Big[(-1-12\omega+4\eta\\
&+48\xi H^2)\dot{B}_k^2+\big[m^2-48\xi H^4-4\eta\dot{H}\\
&-(1-6\omega+12\eta+48\xi \dot{H})H^2\big]B_k^2\Big]\\
&+[-1-12\omega+48\xi(\dot{H}+H^2)]H\dot{B}_kB_k\\
&+2(\eta-3\omega+12\xi H^2)\ddot{B}_kB_k\Big\}\delta_j^i,
\end{aligned}
\end{equation}
where, each component of the density and pressure given by $T_0^{0(k)}=\rho_k$ and $T_j^{i(k)}=-p_k\delta_j^i$ satisfies the consistency relation $\dot{\rho}_k+3H(\rho_k+p_k)=0$, for $k=1-5$, which was carefully verified  for this model.\\

\noindent  In order to analyze a possible slow-roll inflation regime generated with this model,  we consider in the next section  suitable values (or constraints) of model parameters $\eta$, $\omega$ and $\xi$, taking into account that in Eq. (\ref{eq16}) the restriction $\dot{H}\ll m^2_{\rm{eff}}\ll H^2$ must be satisfied.

\subsection{The case $\eta=0$ and $\omega=1/6$}\label{sec4}
\noindent In this section we consider that $\eta=0$, $\omega=1/6$ and the  parameter  $\xi$ take arbitrary values (but considering that the slow-roll conditions and the observational constraints must be satisfied). We can see that, from Eqs. (\ref{eq15}), (\ref{eq21}) and (\ref{eq22})  the background equations reads
\begin{equation}\label{eq23}
\ddot{B}_k+3H\dot{B}_k+[m^2+24\xi H^2(\dot{H}+H^2)]B_k=0,
\end{equation}
\begin{equation}\label{eq24}
3H^2=\frac{3\kappa^2}{2}\Big[(\dot{B}_k^2+m^2B_k^2)+48\xi H^3\dot{B}_kB_k\Big],
\end{equation}
\begin{equation}\label{eq25}
\begin{aligned}
2\dot{H}=&\kappa^2\Big[3(-1+8\xi H^2)\dot{B}_k^2-24 \xi H^2[m^2\\
&+24 \xi H^2(\dot{H}+H^2)]B_k^2\\
&+48 \xi H(\dot{H}-2H^2)\dot{B}_kB_k\Big].
\end{aligned}
\end{equation}
where the equation of motion Eq. (\ref{eq23}) was used to simplify the last expression (Eq. (\ref{eq25})). Note that  in our case the  right hand side of the Eqs. (\ref{eq24}) and (\ref{eq25})  are multiplied by 3, which is due to the number of vector field components.
These equations are similar to those studied in models of inflation with scalar fields coupled to the Gauss-Bonnet invariant  [\cite{carter, guo, hu, lee, kanti, bruck}]. 
Therefore, the slow-roll approximation $\dot{\phi}^2/2\ll V$ for the inflaton field is equivalent (for massive vector fields) to the condition $\dot{B}_k^2\ll m^2B_k^2$. In a similar fashion (replacing $\phi$ by  $B_k$ in the scalar field equations), we can calculate for this model the number of e-folds and the observable quantities such as the power spectra of the scalar and tensor modes, the spectral indices, the tensor-to-scalar ratio and the running spectral indices. For example, considering  the slow-roll conditions (see  [\cite{guo}] and [\cite{lee}])
\begin{equation}\label{eq26}
\begin{aligned}
&\dot{B}_k^2\ll m^2B_k^2, \ \ \ \ \ddot{B}_k\ll 3H\dot{B}_k,\\
&24\xi H\dot{B}_k B_k\ll 1, \ \ \ \ \dot{H}\ll H^2.
\end{aligned}
\end{equation}
the background equations Eqs. (\ref{eq23}), (\ref{eq24}) and (\ref{eq25}), are reduced to
\begin{equation}\label{eq27}
3H\dot{B}_k+m^2B_k+24\xi H^4B_k\simeq 0,
\end{equation}
\begin{equation}\label{eq28}
H^2\simeq \frac{\kappa^2}{2}m^2B_k^2,
\end{equation}
\begin{equation}\label{eq29}
\dot{H}\simeq -\frac{3\kappa^2}{2}(\dot{B}_k^2+32\xi H^3\dot{B}_kB_k).
\end{equation}
The number of e-folds is given by (assuming that  $B_{k(e)}^2$ is negligible compared to $B_{k(i)}^2$)
\begin{equation}\label{efolds}
\begin{aligned}
N=&\int _{t_{i}}^{t_{e}} Hdt\simeq \int _{B_{k(e)}^2}^{B_{k(i)}^2}\frac{3}{4\big[1+6\alpha(\kappa^2B_k^2)^2\big]}d\big(\kappa^2B_k^2\big)\\
&\simeq \frac{1}{4}\sqrt{\frac{3}{2\alpha}}\tan^{-1}{\big[\sqrt{6\alpha}\,\kappa^2 B_{k(i)}^2\big]},
\end{aligned}
\end{equation}
where $\alpha\equiv m^2\xi$ is a dimensionless parameter and we have used the Eqs. (\ref{eq27}) and (\ref{eq28}). Moreover, summation over repeated indices is understood.
\begin{figure}[t]
\centerline{\includegraphics[width=0.7\linewidth,scale=1,angle=-90]{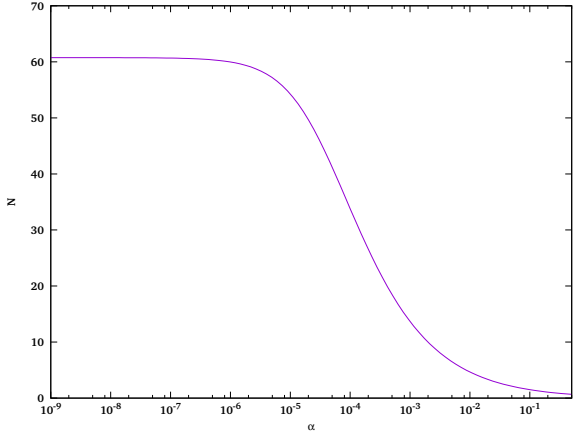}}
\caption{Plot of the e-folding number $N$ versus the parameter $\alpha$, using $B_{k(i)}^2=81\,M_p^2$. \label{fig1}}
\end{figure}
We can see in Fig. \ref{fig1} that the  condition of $N\gtrsim 60$ requires $\alpha\lesssim\alpha_c=10^{-6}$, where $\alpha_c$ is the
value when $\alpha$ becomes nearly constant. In this case, the e-folding number $N$  behaves in a similar way to that shown in Fig. 2 of the reference [\cite{lee}]. \\

\noindent Thereby, all results obtained with  the scalar-Gauss-Bonnet inflation models are reproduced with this proposal (as long as the value of the parameter $\xi$ is according to slow-roll conditions and
the observational constraints), so the vector fields should be considered as a good candidate to drive inflation. But,  it is necessary to realize  a detailed study of stability to guarantee the viability of the model.\\

\subsection{The case $\eta=-2/3$ and $\omega=0$}\label{sec4}
The choice for the model parameters in this section are $\eta=-2/3$ and $\omega=0$. With this selection, the term proportional to $H^2$ in the Eq. (\ref{eq15}) is suppressed and
 the restriction $m^2_{\rm{eff}}\ll H^2$ it is  satisfied. Now, using again Eqs. (\ref{eq15}), (\ref{eq21}) and (\ref{eq22})  the basic equations are given by
\begin{equation}\label{eq30}
\ddot{B}_k+3H\dot{B}_k+\left[m^2+\frac{1}{3}\dot{H}+24\xi H^2(\dot{H}+H^2)\right]B_k=0,
\end{equation}
\begin{equation}\label{eq31}
\begin{aligned}
3H^2=&\frac{3\kappa^2}{2}\Big[(\dot{B}_k^2+m^2B_k^2)+\frac{2}{3}H\dot{B}_kB_k\\
&+H^2B_k^2+48\xi H^3\dot{B}_kB_k\Big],
\end{aligned}
\end{equation}
\begin{equation}\label{eq32}
\begin{aligned}
2\dot{H}=&\kappa^2\Big[\frac{2}{3}(-5+36\xi H^2)\dot{B}_k^2\\
&+\frac{1}{9}\big[3m^2+16\dot{H}-18H^2[-1+12m^2\xi\\
&+4\xi H^2(-1+72\xi(\dot{H}+H^2))]]\big]B_k^2\\
&+2H[1+24\xi(\dot{H}-2H^2)]\dot{B}_kB_k\Big].
\end{aligned}
\end{equation}
Using the slow-roll conditions given by the Eq. (\ref{eq26}), the  background equations Eqs. (\ref{eq30}), (\ref{eq31}), and (\ref{eq32}) take the following form
\begin{equation}\label{eq33}
3H\dot{B}_k+m^2B_k+24\xi H^4B_k\simeq 0,
\end{equation}
\begin{equation}\label{eq34}
H^2\simeq \frac{\kappa^2}{2}(m^2+H^2)B_k^2,
\end{equation}
\begin{equation}\label{eq35}
\dot{H}\simeq -\kappa^2\left(\frac{5}{3}\dot{B}_k^2-H\dot{B}_kB_k\right).
\end{equation}
But, taking into account that $m^2\ll H^2$ in Eq. (\ref{eq34}), we have that the vector field is reduced to a constant (the fields $B_k$ are ``frozen''): 
\begin{equation}\label{eq36}
B_k^2\simeq 2M_p^2,
\end{equation}
Therefore,  the energy density $\rho\simeq (3/2)\kappa^2H^2B_k^2$   associate to  the vector field  remains almost constant and the potential $V=-(1/2)m^2B_k^2=const$ could drive the quasi de Sitter expansion analogous to the scalar field (chaotic Inflation) [\cite{linde2, golovnev}],  hence that slow-roll takes place and inflation occurs. In this context, the inflation period does not finish (eternal inflation) and therefore, the number of e-folds $N$ is not considered in this case.\\

\noindent In general, considering that the model parameters satisfies the constraint $2+3\eta-12\omega=0$,  again the term proportional to $H^2$ in the Eq. (\ref{eq15}) is removed and the restriction $m^2_{\rm{eff}}\ll H^2$ it is  satisfied. Hence, in a similar way to the last case, the vector field is reduced to a constant:
\begin{equation}\label{eq37}
B_k^2\simeq \left(\frac{2}{1-6\omega}\right)M_p^2,
\end{equation}
where $1-6\omega\neq 0$, and the non-minimal coupling with $\omega=1/6$ is forbidden in this context.

\section{Conclusions}\label{sec_concs}
\noindent In this work we have considered slow-roll inflation for a  vector-tensor model  with  massive vector fields non-minimally coupled to gravity.  In addition,  the model under consideration has arbitrary parameters for each geometrical coupling (see Eq. (\ref{eq1})). Using a spatially flat FRW type universe and  general vector fields with temporal and spatial components, the general expressions for equation of motion and the  components of the energy momentum tensor were obtained (see Eqs. (\ref{eq15}),  (\ref{eq17}) and (\ref{eq18})).
In this scenario, the isotropy of expansion  is guaranteed considering a triplet of orthogonal vector fields, but the effective mass of the vector field is of the  order of the Hubble scale  and  the inflationary regime is difficult to realize with this model (see Eq. (\ref{eq16})). However, for suitable values (or constraints) of model parameters it was possible to overcome this issue. For the case $\eta=0$ and $\omega=1/6$, the  model it behaves in a  similar way to those  models of inflation with scalar fields coupled to the Gauss-Bonnet invariant (see Eqs. (\ref{eq27}), (\ref{eq28}), (\ref{eq29}) and Fig. \ref{fig1}). The analysis with $\eta=-2/3$ and $\omega=0$ was performed and with this choice, it is shown that the vector field  is reduced to a constant (see Eq. (\ref{eq36})), and the vector field drives a quasi de Sitter expansion,
(analogous to the scalar field), hence that slow-roll takes place and inflation occurs. On the other hand, for the constraint $2+3\eta-12\omega=0$, again the vector field is constant 
and the non-minimal coupling with $\omega=1/6$ is forbidden (see Eq. (\ref{eq37})). This model could be generalized taking into account  an arbitrary potential 
 $V(A^2)$   and an arbitrary non-minimal couplings $\eta(A^2)$, $\omega(A^2)$ and $\xi(A^2)$, where $A^2\equiv A_{\mu}A^{\mu}$. Finally, from the above analysis we can say that  the vector fields non-minimally coupled to gravity can be considered as a good candidate to drive inflation. However, it is necessary to perform  a detailed study of stability to guarantee the viability of the model.



\nocite{*}
\bibliographystyle{spr-mp-nameyear-cnd}

\begin{thebibliography}{99}

\bibitem[Albrecht et al. (1982)]{albrecht} Albrecht A.,   Steinhardt P. J.: Phys. Rev. Lett. \textbf{48}, 1220 (1982).
\bibitem[B\"ohmer et al. (2007)]{harko} B\"ohmer C. G.,  Harko T.: Eur. Phys. J. C \textbf{50}, 423 (2007). arXiv:gr-qc/0701029
\bibitem[Bruck et al. (2016)]{bruck}  Bruck C. van de, Dimopoulos K.,  Longden Ch.: Phys. Rev. D \textbf{94}, no.2, 023506 (2016). arXiv:1605.06350 [astro-ph.CO]
\bibitem[Burd et al. (1991)]{burd} Burd A. B.,  Lidsey J. E.: Nucl. Phys. B \textbf{351}, 679 (1991).
\bibitem[Chiba (2008)]{chiba} Chiba T.: JCAP \textbf{0808}, 004 (2008). arXiv:0805.4660 [gr-qc]
\bibitem[Darabi et al. (2014)]{darabi}  Darabi F., Parsiya A.: Int. J. Mod. Phys. D \textbf{23}, no.08, 1450069 (2014). arXiv:1401.1280 [gr-qc]
\bibitem[Ford (1989)]{ford} Ford L. H.: Phys. Rev. D \textbf{40},  967 (1989).
\bibitem[Germani et al. (2009)]{germani} Germani C.,  Kehagias A.:  JCAP \textbf{0903}, 028 (2009).  arXiv:0902.3667 [astro-ph.CO]
\bibitem[Golovnev (2008)]{golovnev} Golovnev A., Mukhanov V., Vanchurin V.: JCAP \textbf{0806}, 009 (2008).  arXiv:0802.2068 [astro-ph]
\bibitem[Granda et al. (2014)]{granda} Granda L. N., Jimenez D. F.: Phys. Rev. D \textbf{90},  no.12, 123512 (2014).  arXiv:1411.4203 [gr-qc]
\bibitem[Guo et al. (2010)]{guo} Guo Z-K.,  Schwarz D. J.: Phys. Rev. D \textbf{81}, 123520 (2010). arXiv:1001.1897 [hep-th]
\bibitem[Guth (1981)]{guth}  Guth A.: Phys. Rev. D \textbf{23}, 347 (1981).
\bibitem[Guth et al.  (1982)]{guth2} Guth A. H.,  Pi S. Y.: Phys. Rev. Lett. \textbf{49}, 1110 (1982).
\bibitem[Hawking (1982)]{hawking} Hawking S. W.: Phys. Lett. B \textbf{115}, 295 (1982).
\bibitem[Himmentoglu et al. (2009)]{peloso} Himmetoglu B., Contaldi C. R.,  Peloso M.: Phys. Rev. Lett. \textbf{102}, 111301 (2009). arXiv:0809.2779 [astro-ph]
\bibitem [Jiang et al. (2013)]{hu}  Jiang P-X., Hu J-W.,  Guo Z-K.: Phys. Rev. D \textbf{88}, 123508 (2013).  arXiv:1310.5579 [hep-th]
\bibitem[Jimenez et al. (2008)]{maroto1} Jimenez J. B., Maroto A. L.: Phys. Rev. D \textbf{78}, 063005 (2008).  arXiv:0801.1486 [astro-ph]
\bibitem[Jimenez et al. (2009)]{maroto2} Jimenez J. B., Lazkoz R.,  Maroto A. L.: \textit{Phys. Rev. D} \textbf{80}, 023004 (2009).  arXiv:0904.0433 [astro-ph.CO] 
\bibitem[Kanti et al. (2015)]{kanti} Kanti P.,  Gannouji R.,  Dadhich N.: Phys. Rev. D \textbf{92}, no.4, 041302 (2015).  arXiv:1503.01579 [hep-th]
\bibitem[Kiselev (2004)]{kiselev} Kiselev V. V.: Class. Quant. Grav. \textbf{21}, 3323 (2004).  gr-qc/0402095
\bibitem[Koh (2011)]{koh}  Koh S.: Int. J. Mod. Phys. Conf. Ser. \textbf{01}, 120 (2011).  arXiv:0902.3904 [hep-th]
\bibitem[Koh et al. (2012)]{koh2} Koh S.,  Hu B.:  J. Korean Phys. Soc. \textbf{60}, 1983 (2012).  arXiv:0901.0429 [hep-th]
\bibitem[Koh et al. (2014)]{lee}  Koh S., Lee B-H.,  Lee W.,  Tumurtushaa G.: Phys. Rev. D \textbf{90}, no.6, 063527 (2014).  arXiv:1404.6096 [gr-qc]
\bibitem[Koivisto et al. (2008)]{mota} Koivisto T. S.,  Mota D. F.:  JCAP  \textbf{0808}, 021 (2008).  arXiv:0805.4229 [astro-ph]
\bibitem[Liddle et al. (2000)]{liddle}  Liddle A. R., D. H. Lyth D. H.: ``Cosmological inflation and large-scale structure'', Cambridge
University Press (2000).
\bibitem[Lidsey et al. (1997)]{lidsey}  Lidsey J. E., Liddle A. R., Kolb E. W.,  Copeland E. J., Barreiro T.,  Abney M.: Rev. Mod. Phys. \textbf{69}, 373 (1997).  astro-ph/9508078 
\bibitem[Linde (1982)]{linde} Linde A. D.:  Phys. Lett. B \textbf{108},  389 (1982).
\bibitem[Linde (1990)]{linde2} Linde A.: ``Particle Physics and Inflationary Cosmology'', Harwood, Chur (1990).
\bibitem[Lyth et al.  (1999)]{lyth}  Lyth D. H., Riotto A.: Phys. Rept. \textbf{314},  1 (1999).  hep-ph/9807278 
\bibitem[Maleknejad et al. (2013)] {soda} Maleknejad A.,  Sheikh-Jabbari M. M.,  Soda J.: Phys. Rept. \textbf{528}, 161 (2013).  arXiv:1212.2921 [hep-th]
\bibitem[Mukhanov et al. (1981)]{chibisov} Mukhanov V. F., Chibisov G. V.: JETP Lett. \textbf{33},  532 (1981).
\bibitem[Neupane et al. (2006)]{carter} Neupane I. P.,   Carter B. M. N.: JCAP \textbf{0606}, 004 (2006).  hep-th/0512262 
\bibitem[Nojiri et al. (2005)]{odintsov}  Nojiri S.,  Odintsov  S. D.,  Sasaki M.: Phys. Rev. D \textbf{71}, 123509 (2005).  hep-th/0504052 
\bibitem[Oliveros et al. (2015)]{oliveros}  Oliveros A.,  Solis  E. L.,  Acero M. A.: Mod. Phys. Lett. A \textbf{31}, no.01, 1650009 (2015). arXiv:1510.04673
\bibitem[Setare et al. (2013)]{setare} Setare M. R., Kamali V.: Phys. Lett. B \textbf{726}, 56 (2013).  arXiv:1309.2452 [gr-qc]
\bibitem[Wei et al. (2006)]{wei} Wei H.,   Cai R-G.: Phys. Rev. D \textbf{73}, 083002 (2006).  astro-ph/0603052












 






\
\end{thebibliography}

\end{document}